\documentclass[prl,twocolumn,showpacs,preprintnumbers,amsmath,amssymb]{revtex4}
\usepackage{txfonts}
\usepackage{epsfig,amsmath,amssymb} 
\usepackage[all]{xy}
\usepackage{ifpdf} 
\usepackage{graphicx}   
\usepackage{amsfonts}
\usepackage{makeidx}  
\usepackage{epigraph}
\usepackage{etoolbox}
\makeatletter
\patchcmd{\epigraph}{\@epitext{#1}}{\itshape\@epitext{#1}}{}{}
\makeatother

\newcommand{\qed}{\hfill \mbox{\raggedright \rule{.07in}{.1in}}}

\newcommand{\ket}[1]{\left | #1 \right\rangle}

\newcommand{\braket}[2]{\left\langle #1|#2\right\rangle}

\newcommand{\qq}[1]{``#1"}

\begin{document}

\title{Quantum-gravity effects could in principle be witnessed in neutrino-like oscillations} 

\author{C. Marletto $^{a, b, c}$, V. Vedral $^{a, b, c}$, and D. Deutsch $^{a}$
	\\ {\small $^{a}$ Clarendon Laboratory, Department of Physics, University of Oxford} 
	\\{\small $^{b}$Centre for Quantum Technologies and Department of Physics, National University of Singapore}
		\\{\small $^{c}$Institute for the Scientific Interchange, Turin, Italy}}

\date{April 2018}

\begin{abstract}
Two of us (CM and VV \cite{MAVE}) recently showed how the quantum character of a physical system, in particular the gravitational field, can in principle be witnessed without directly measuring observables of that system, solely by its ability to mediate entanglement between two other systems. Here we propose a variant of that scheme, where the entanglement is again generated via gravitational interaction, but now between two particles both at sharp locations (very close to each other) but each in a superposition of two different masses. We discuss an in-principle example using two hypothetical massive, neutral, weakly-interacting particles generated in a superposition of different masses. The key property of such particles would be that, like neutrinos, they are affected only by weak nuclear interactions and gravity. \noindent  \bigskip

\end{abstract}

\maketitle

The domain where quantum theory and theories of gravitation intersect is traditionally probed by considering experiments using a single mass in a superposition of two different locations, as in Feynman's thought experiment \cite{FEY}. Two recent schemes have been proposed extending this tradition \cite{MAVE, SOUG}. Their key innovation is to use the gravitational field as a {\sl mediator} to produce entanglement between {\sl two } masses, each prepared in a quantum superposition of two different locations. As explained in \cite{MAVE}, the entanglement generated between the two masses (in the path degree of freedom) is an indirect witness of the quantisation of the gravitational field, in that it requires the field to have at least two variables that cannot be simultaneously measured. Here we discuss a hypothetical variant of that scheme, where the gravitational interaction takes place between two particles that are each in a superposition of two different {\sl masses}. 

Any particle that is in a superposition of two non-degenerate eigenstates of its Hamiltonian is in a superposition of two different generalised masses. Typically, gravitational effects on, or of, particles in such superpositions are too weak to be detected, and are swamped by decoherence from other sources. But not, for example, for neutrinos emitted in the phenomenon of double-beta decay, since neutrinos interact exclusively via gravity and the weak interaction. The latter has an extremely short range (about $10^{-18}$ m) and is therefore negligible at the scale of the size $d$ of the detector, which we envisage as something like a nucleus ($10^{-15}$ m).

Neutrinos are neutral particles produced in $\beta$-decay \cite{GIU}, for instance the decay of a neutron into a proton, an electron and an electron-antineutrino. (For brevity we shall omit \qq{€˜anti} in what follows.) In the Standard Model, neutrinos were originally assumed to be massless; but the phenomenon of neutrino oscillations reveals that this is false \cite{GIU}. In fact a neutrino can exist in at least three different flavours, each eigenstate of flavour being a superposition of different states of mass. 

Here we shall calculate the effect of the gravitational interaction between two hypothetical neutrino-like particles. As we shall see, the interaction affects the period of the oscillations, which could in principle be used to demonstrate, or witness, gravitationally-mediated entanglement between the two particles. 

Two observables are relevant for the oscillations of a single neutrino-like particle: one is its Hamiltonian $H$ (depending only on the internal degrees of freedom), with three eigenstates $\ket{m_i}$, which we label by the value of the rest mass, $m_i$. The other observable, which does not commute with $H$, is the flavour, with three eigenstates $\ket{\nu_i}$ in the flavour subspace.  

We first analyse the oscillation of an isolated particle; and then we will see how the effect of the mutual interaction with another like particle changes the phases, thus providing the desired witness of entanglement.  We shall work in the particle's rest-frame and assume that, like the real neutrino, it is created initially in a flavour eigenstate $\nu_1$, which, for simplicity, we shall assume is a superposition of two of the eigenstates of the Hamiltonian:
\begin{equation}
\ket{\nu_1}=\cos (\theta)\ket{m_1}+\sin(\theta)\ket{m_2}
\end{equation}
where $\theta$ is the mixing angle -- a fixed parameter which, as for neutrinos, would depend on the physics of weak interactions \cite{GIU}. 

Now, suppose that the created particle evolves freely under its Hamiltonian $H$; and then at time $t$ its flavour is measured. The state of the particle just before the measurement is:
\begin{equation}
\ket{\psi(t)}=\cos (\theta)e^{im_1c^2\frac{t}{\hbar}}\ket{m_1}+\sin(\theta)e^{im_2c^2\frac{t}{\hbar}}\ket{m_2}.
\end{equation}
Thus the \qq{survival} probability for particle still to be in the initial flavour eigenstate $\ket{\nu_1}$ when the detection happens is $$P_s=|\braket{\nu_1}{\psi(t)}|^2=1-\sin^2(2\theta)\sin^2\left (\Delta mc^2\frac{t}{2\hbar}\right ),$$where $\Delta m= m_2-m_1$.

In our thought experiment, we are interested in the effects of gravitational interactions between two of these particles that are produced very close together, simultaneously. 
Since the particles interact gravitationally, the phases of the flavour components of their states are further modified by the gravitational potential, at each of them, due to the other. As proved in \cite{MAVE}, this causes entanglement which is a witness of the quantum nature of the gravitational field.

So, consider two such particles separated by a distance $d$, where $d$ is about the size of a nucleus.
Initially they are both in the same state $\ket{\nu_1}$, i.e., their state is $\ket{\nu_1}\ket{\nu_1}$. Let us calculate the phase shift due to their mutual gravitational interaction. 

As a result of the interaction, the initial state of the two particles $\ket{\nu_1}\ket{\nu_1}$ evolves into the state $$\ket{\phi(t)}=\sum_{i,j}\alpha_{ij}\ket{m_i }\ket{m_j}$$ where 
$$\alpha_{ii}=\exp{ \left \{- i\left (2m_ic^2 +\frac{Gm_i^2}{d}\right )\frac{t}{\hbar}\right \}}$$ and $$\alpha_{12}=\alpha_{21}=\exp{\left \{-i\left ((m_1+m_2)c^2+\frac{Gm_1m_2}{d}\right )\frac{t}{\hbar}\right \}}\;.$$ Thus, the degree of entanglement varies with $t$.  

The probability that a detector acting at time $t$ will detect the flavour eigenstate
 $\ket{\nu_1}$ is now modified by the gravitational interaction: 
 $$|\braket{\nu_1}{\phi(t)}|^2=1-\sin^2(2\theta)\sin^2\left\{\left(\frac{\Delta m c^2}{2}+\frac{Gm_1\Delta m}{d}\right)\frac{ t}{\hbar}\right\}.$$
 
The additional gravitationally induced phase $\Phi_G=G\frac{m(\Delta m)}{d\hbar }t$ would be extremely small for neutrinos, because of their tiny masses (of the same order as $\Delta m\approx 10^{-38}kg$). 
But we can imagine a hypothetical neutrino-like particle with much larger masses. Suppose for example that they are of the order of the Planck mass $m=10^{-8} kg$. (Note that the ratio $\frac{\Phi_G}{\Phi}$, where $\Phi=\frac{\Delta mc^2t}{\hbar}$, depends on $m_1+m_2$, but not on $\Delta m$). The particle could then not be created by a nuclear decay, since nuclei aren't that massive. It would have to come from some much more energetic event about whose nature we need not speculate, except to say that it must happen at a location fixed to an accuracy $d$. 

Given our assumption that, like neutrinos, the particle interacts solely via 
the gravitational and weak nuclear forces, that still-tiny, gravitationally
induced phase change $\Phi_G$ could be detectable in principle. For 
suppose that the source and the detector are a distance $L$ apart, so that 
detection occurs at a time $\frac{L}{v}$ (in the laboratory frame) after 
emission, where $v$ is the speed of the particle. The detection probability 
as a function of $L$ will be periodic with wavelength $\lambda=2\pi \frac 
{c\gamma}{\omega}$ where $\omega=\frac{\Delta m c^2}{2\hbar}$, and $
\gamma=\frac{1}{\sqrt{1-\frac{v^2}{c^2}}}$ is the Lorentz factor. If we 
assume that the detector has the same size $d$ as the source, the 
condition for this spatial variation to be detectable is $\lambda>d$. We 
also assume that the detector can distinguish a 2-particle hit from a 1-
particle hit. The latter will typically be from one member of a pair whose 
other member travelled in a different direction, so that the detected 
particle was subjected to a negligible mutual gravitational interaction on its 
journey, leading to a negligible gravitational phase shift. Hence at distances $L$ where $\frac{\Phi_G}{\Phi}\frac{
\Delta m c^2L}{\gamma \hbar v}=(n+\frac{1}{2})\pi$ for integer $n$, the oscillations of affected 
and unaffected particles will be out of phase, so that only single- (or 
alternately only double-) particle hits will be detected. 

As the particles travel to the detector, their wave packets will spread. If the initial spread in position is $\delta$, the spread in speed can be as little as $\frac{\hbar}{\gamma m \delta}$, so the final spread in position will be about $\delta+\frac{\hbar L}{\gamma mv\delta}$, which can be as low as $2\sqrt{\frac{\hbar L}{\gamma mv}}$ so we must have $2\sqrt{\frac{\hbar L}{\gamma mv}}<\frac{d}{2}$.

These conditions are met for a range of values of $v$ and $\Delta m$ -- for instance $d\approx 10^{-15}m$, $\gamma \approx10^4$, $\frac{L}{v} \approx 10^{-1}$s, and $\Delta m\approx 10^{-25}kg$. 

The different gravitational potentials due to other masses, at the two paths that are $d$ apart, would also contribute to the phase. We may suppose that the experiment is done in free fall, far from all large masses. For a mass $M$ at a distance $R$ not to swamp the effect we are measuring, we must have $\frac{dGM\Delta m }{R^2} < \frac{Gm\Delta m}{d}$, i.e. $\frac{d^2}{R^2} < \frac{m}{M}$. Particles with the parameters we have considered easily satisfy this condition if the experiment is feasible in Earth orbit (with, say, $R\approx 10^{7}$m, and the mass of the Earth about $6\times10^{24}$kg).

\textit{Acknowledgments}: 
The Authors thank Carlo Giunti for helpful discussions about neutrino physics. CM's research was supported by the Templeton World Charity Foundation and the Eutopia Foundation. VV thanks the Oxford Martin School, the John Templeton Foundation, the EPSRC (UK). This research is also supported by the National Research Foundation, Prime Minister's Office, Singapore, under its Competitive Research Programme (CRP Award No. NRF- CRP14-2014-02) and administered by Centre for Quantum Technologies, National University of Singapore.


\begin{thebibliography}{1}
\bibitem{FEY} R. Feynman, Chapel Hill Conference Proceedings, 1957.
\bibitem{MAVE} C. Marletto, V. Vedral, Gravitationally-induced entanglement between two massive particles is sufficient evidence of quantum effects in gravity, Phys. Rev. Lett. 119, 2017.
\bibitem{SOUG} S. Bose et al., Spin Entanglement Witness for Quantum Gravity, Phys. Rev. Lett. 119, 2017.
\bibitem{GIU} C. Giunti, C.W. Kim, Fundamentals of Neutrino Physics and Astrophysics, Oxford University Press, (2007).







		

	%
	
	
\end{thebibliography}
\end{document}